# High-efficiency entanglement of microwave fields in cavity opto-magnomechanical systems

Ke Di[1], Shuai Tan[1], Liyong Wang[2,*], Anyu Cheng[1], Xi Wang[1], Yu Liu[1], Jiajia Du[1,*]

1 Chongqing University of Post and Telecommunications, Chongqing 400065, China
2 Department of Applied Physics, Wuhan University of Science and Technology, Wuhan 430081, China

E-mail: **wangliyong@wust.edu.cn** and **dujj@cqupt.edu.cn**

January 2023

Abstract. We demonstrate a scheme to realize high-efficiency entanglement of two microwave fields in a dual opto-magnomechanical system. The magnon mode simultaneously couples with the microwave cavity mode and phonon mode via magnetic dipole interaction and magnetostrictive interaction, respectively. Meanwhile, the phonon mode couples with the optical cavity mode via radiation pressure. Each magnon mode and optical cavity mode adopts a strong red detuning driving field to activate the beam splitter interaction. Therefore, the entangled state generated by the injected two-mode squeezed light in optical cavities can be eventually transferred into two microwave cavities. A stationary entanglement $E_{a_1 a_2}$ = 0.54 is obtained when the input two-mode squeezed optical field has a squeezing parameter r=1. The entanglement $E_{a_1 a_2}$ increases as the squeezing parameter r increases, and it shows the flexible tunability of the system. Meanwhile, the entanglement survives up to an environmental temperature about 385 mK, which shows high robustness of the scheme. The proposed scheme provides a new mechanism to generate entangled microwave fields via magnons, which enables the degree of the prepared microwave entanglement to a more massive scale. Our result is useful for applications which require high entanglement of microwave fields like quantum radar, quantum navigation, quantum teleportation, quantum wireless fidelity (Wi-Fi) network, etc.

## 1. Introduction

Entangled states have been widely investigated for decades since it is an essential resource in fields of quantum computation [1], quantum key distribution [2–4], quantum teleportation [5–7], quantum network construction [8, 9] and fundamental physics [10, 11]. The entangled state of light field exists in two forms due to the wave-particle duality. One is the entangled photon pair in discrete variables (DV), and the other is the two-mode squeezed state in continuous variable (CV). In particular, the light field of CV has the characteristics of superior broadband, excellent stabilization and high operating efficiency. Generally, entangled optical fields in CV are prepared by second-order or third-order nonlinear crystal by parametric down-conversion process [12–15].

Compared to optical fields, there are few methods to prepare entangled microwave fields of CV. The nonlinear Josephson circuit parametric amplifier [16–18], circuit quantum electrodynamics (QED) [19, 20] and optomechanics [21, 22] systems are traditionally used to generate entangled states of high quality microwave fields. In 2020, preparing steady-state entanglement of two microwave fields based on nonlinear magnetostrictive effects in a cavity magnomechanical system is reported [23], and it is easy to satisfy the resolved-sideband condition to cool the mechanical oscillator to ground-state since the linewidth of the magnon is much smaller than the phonon frequency [24].

In recent years, magnons in cavity magnomechanical systems has been demonstrated the great potential due to their excellent coupling properties. Cavity magnomechanical system is useful for studying quantum states at macroscopic scales [25–29]. As a ferromagnetic material, Yttrium iron garnet (YIG) is a key component of the cavity magnomechanical system [25, 26], and it has high spin density and low dissipation rate [28, 30]. The magnon is an embodiment of the collective excitation for spin waves inside the YIG, and it can strongly couple with the microwave cavity mode via magnetic dipole interaction [25, 31]. In addition, the magnon mode can also couple with the phonon mode by magnetostriction-induced deformation of YIG crystal. Recently, it has been demonstrated that the phonon mode can be successfully coupled to the optical cavity mode via radiation pressure [32, 33]. It provides a route for communication of entangled state between microwave field and optical field. Due to the strong coupling characteristics of magnon [34–36], bipartite or tripartite entanglement in a cavity magnomechanical system is feasible [31, 37]. For example, Magnons can be used to generate magnomechanically induced transparency (MMIT) phenomena, which have been observed experimentally [30, 38, 39]. Lately, multi-channel MMIT phenomena have been realized by coupling magnon modes with multiple different physical subsystems [40, 41]. In 2020, M. Yu et al. proposed a scheme for entangling two microwave fields in a cavity magnomechanical system. The entanglement of two microwave fields reaches up to $E_N$=0.18, and the entanglement survives at an environmental temperature about 140 mK [23]. Since then, How to further enhance the microwave entanglement $E_N$ and the robustness temperature T is an open problem.

In this paper, we propose a scheme to generate high-efficiency stationary entanglement of two microwave fields in a dual opto-magnomechanical system which consists of two optical cavities and two microwave cavities. Two YIG crystals are embedded in each microwave cavity under an uniform bias magnetic field. A microwave driving field is applied in the perpendicular direction to activate the phonon mode. Meanwhile, the magnons couple with the microwave fields via magnetic dipole interaction, and the phonons couple with the optical cavity modes via radiation pressure. A two-mode squeezed optical field is injected into two optical cavities which makes two optical cavities quantum correlated. Each optical cavity is driven by a red-detuned laser field to activate the optomechanical beam splitter interaction. Therefore, the entanglement between two optical cavities is transferred to two phonon modes. Simultaneously, two magnon modes are driven by strong red detuning microwave fields to activate magnomechanical beam splitter interactions. Then the entanglement of two phonon modes is further transferred to two magnon modes. Furthermore, two microwave cavities are entangled due to the magnon-microwave beam splitter interaction. As a consequence, quantum state transfer is realized from photons to phonons, then to magnons, and finally to microwaves. The proposed scheme for preparing stationary entanglement of two microwave fields is high efficiency ($E_{a1a2}$ = 0.54) and robustness to the environmental temperature (T=385 mK), and it will be useful for a variety of applications like quantum radar [40, 41], quantum teleportation [4, 5, 6], quantum network construction [7, 8], fundamental physics [9, 10], quantum wireless fidelity (Wi-Fi) network, etc.

## 2. Theoretical Model

Fig.1 shows the dual-cavity opto-magnomechanical system. Each cavity opto- magnomechanical subsystem consists of a microwave cavity mode, a magnon mode, a phonon mode and an optical cavity mode. The magnon and phonon modes are constructed by YIG crystal (a 5 × 2 × 100 μm³ YIG cuboid) with a micro-bridge structure [30, 31]. The YIG crystal is placed in the microwave cavity and in a bias magnetic field. The optical cavity consists of two highly reflective mirrors. The right cavity mirror attached to the surface of a YIG micro-bridge. An uniform bias magnetic field and a microwave field interact with the YIG crystal to activate the magnon mode. The microwave field is injected into the microwave cavity from the right side of the system. The magnon mode couples with the microwave cavity mode and the phonon mode through magnetic dipole interaction and nonlinear magnetostrictive interaction [34, 25]. Furthermore, the phonon mode couples with the optical cavity mode through radiation pressure [31, 30]. The strong red detuning microwave field and optical field are used to drive the magnon mode and optical cavity mode, respectively [29]. The red detuning field here can cool the mechanical motion and increase the magnomechanical (optomechanical) coupling strength, thus the magnon-phonon-photon state-swap interaction is activated [23, 29].

The Hamiltonian for the dual-cavity opto-magnomechanical system can be written as [23, 42]:

$$\begin{aligned} H/\hbar = \sum_{j=1,2} &\{\omega_{a_j} a_j^+ a_j + \omega_{m_j} m_j^+ m_j + \omega_{c_j} c_j^+ c_j + \omega_{b_j} b_j^+ b_j \\ &+ g_{a_j}(a_j^+ m_j + a_j m_j^+) + g_{m_j} m_j^+ m_j (b_j^+ + b_j) + g_{c_j} c_j^+ c_j (b_j^+ + b_j) \\ &+ i\Omega_j(m_j^+ e^{-i\omega_{o_j} t} - m_j e^{i\omega_{o_j} t}) + iE_j(c_j^+ e^{-i\omega_{L_j} t} - c_j e^{i\omega_{L_j} t})\}. \end{aligned} \quad (1)$$

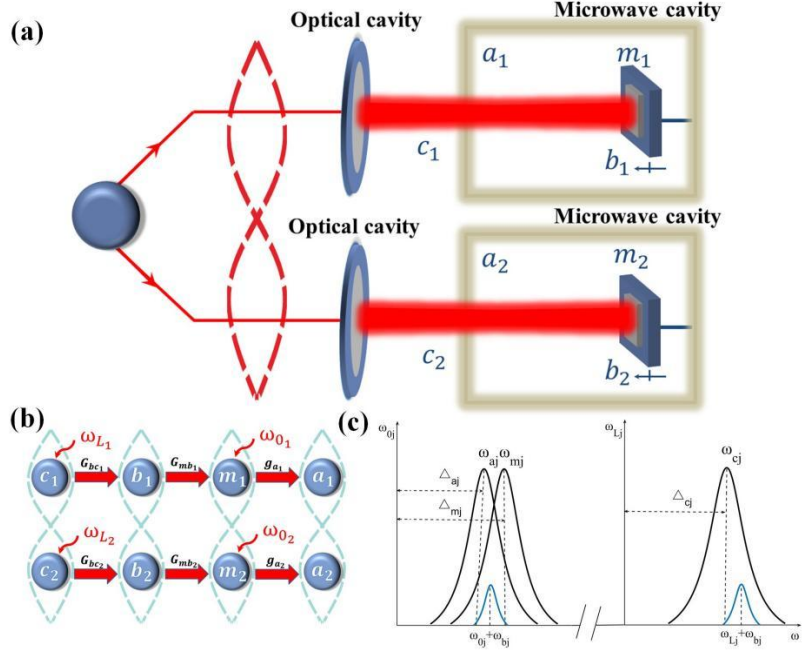

Figure. 1. Scheme diagram of dual-cavity opto-magnomechanical system. (a) Model diagram of the system to realize stationary entanglement of two microwave fields. Two YIG micro-bridges are embedded in two microwave cavities, respectively. High- reflection mirrors are attached to the left side of the YIG crystals, which are used to construct the optical cavities. (b) Scheme diagram of entangled state transfer process. Red arrows denote state-swap interactions. (c) The frequency diagram. j = 1, 2 denote the first subsystem and the second subsystem, respectively. The resonant frequency of the microwave field (optical cavity) is $\omega_{aj}$ ($\omega_{cj}$). The magnon mode with frequency $\omega_{mj}$ is driven by a microwave field with frequency $\omega_{0j}$. The optical cavity is driven by a two-mode squeezed optical field with frequency $\omega_{Lj}$. Both the optical cavity modes and the magnon modes are driven by strong red detuning fields. The mechanical motion triggered by two driving fields generates the Skotos sideband ($\omega_0(L)_j − \omega_{bj}$) and the anti-Skotos sideband ($\omega_0(L)_j + \omega_{bj}$) excitations. Stationary entanglement of two microwave fields can be realized when the optical cavity mode, the magnon mode and the microwave cavity modes in each opto-magnomechanical subsystems resonate with the anti-Skotos sidebands simultaneously.

The first four terms in Eq. (1) describe the energies of four different cavity modes. $a_j$, $m_j$, $c_j$ and $b_j$ ($a_j^+$, $m_j^+$, $c_j^+$ and $b_j^+$) are the annihilation (creation) operators for the microwave cavity mode, the magnon mode, the optical cavity mode and the phonon mode, respectively. It satisfies $[O_j, O_j^+] = 1$ (O = a, m, c, b) and j = 1, 2. $\omega_{Oj}$ to different resonant frequencies of four cavity modes. The frequency of the magnon is $\omega_{mj} = \gamma H_j$. $\gamma / 2\pi = 28$GHz/T is the gyromagnetic ratio and $H_j$ is the external bias magnetic field intensity [23]. The fifth, sixth and seventh terms in Eq.(1) denote the interaction terms of the magnon and microwave cavity modes, the magnon and phonon modes, as well as the phonon and optical cavity modes, respectively. g$_{aj}$ is the magnon-microwave coupling strength, which is larger than the decay rate of the microwave cavity and magnon, i.e., g$_{aj}$ > κ$_{aj}$ , κ$_{mj}$ [33, 43]. κ$_{aj}$ (κ$_{mj}$ ) is decay rate of the microwave cavity (magnon) mode. g$_{mj}$ and g$_{cj}$ are the magnon-phonon coupling rate and phonon-optical cavity coupling rate, respectively. The eighth and ninth terms in Eq.(1) denote the driving fields for the magnon mode and optical cavity mode, respectively. The Rabi frequency $\Omega_j = \frac{\sqrt{5}}{4}\gamma\sqrt{N_{sj}}H_{dj}$ denotes the coupling strength of the magnon mode, where $N_{sj} = \rho V_j$ denotes the total spin number of the ferrimagnet and H$_{dj}$ is the amplitude of the drive magnetic field [23]. ρ = 4.22 × 10$^{27}$ m$^{-3}$ and V$_j$ = 5 × 2 × 100 μm$^3$ are the density and volume of the YIG cuboid. $E_j = \sqrt{2\kappa_{cj}P_{Lj}/(\hbar\omega_{Lj})}$ is the coupling strength of the optical cavity, where κ$_{cj}$ is the decay rate of the optical cavity mode. P$_{Lj}$ = 0.64 mW and ω$_{Lj}$ are the power and frequency of the laser with wavelength of 1550 nm [30].

The quantum Langevin equation for different modes can be obtained by solving equation $\dot{O}_j = \frac{1}{i\hbar}[O_j, H]$. Introduce the input noise terms [44, 23], we get:

$$\begin{aligned}
\dot{a}_j &= -i\Delta_{a_j}a_j - \kappa_{a_j}a_j - ig_{a_j}m_j + \sqrt{2\kappa_{a_j}}a_j^{in}, \\
\dot{m}_j &= -i\Delta_{m_j}m_j - \kappa_{m_j}m_j - ig_{a_j}a_j - ig_{m_j}m_j(b_j^+ + b_j) + \Omega_j + \sqrt{2\kappa_{m_j}}m_j^{in}, \\
\dot{b}_j &= -i\omega_{b_j} - \gamma_j b_j - ig_{m_j}m_j^+ m_j + ig_{c_j}c_j^+ c_j + \sqrt{2\gamma_j}b_j^{in}, \\
\dot{c}_j &= -i\Delta_{c_j}c_j - \kappa_{c_j}c_j + ig_{c_j}c_j(b_j^+ + b_j) + E_j + \sqrt{2\kappa_{c_j}}c_j^{in},
\end{aligned} \quad (2)$$

where $\Delta_n = \omega_n - \omega_{oj}$ ($n = a_j, m_j$), $\Delta_{cj} = \omega_{cj} - \omega_{Lj}$. γ$_{bj}$ is the the mechanical damping rate. $a_j^{in}$, $m_j^{in}$, $b_j^{in}$ and $c_j^{in}$ are the input noise operators corresponding to different modes. The mean value of the input noise $K_j^{in}$ (K = a, m, b) is zero. The correlation functions of the input noises can be described by [45]:

$$\begin{aligned}
\langle K_j^{in}(t)K_j^{in+}(t)\rangle &= (N_{K_j} + 1)\delta(t-t), \\
\langle K_j^{in+}(t)K_j^{in}(t)\rangle &= (N_{K_j})\delta(t-t),
\end{aligned} \quad (3)$$

where $N_{Kj} = \left[\exp(\frac{\hbar\omega K_j}{\kappa_B T}-1)\right]^{-1}$ is mean thermal excitation number of each cavity mode. k$_B$ is the Boltzmann constant and T is the bath temperature. As shown in Fig. 1, the two-mode vacuum squeezed light field is injected into different optical cavities to make the noise operators of two optical cavities quantum correlated. The input noise correlation function for the two-mode squeezed light field can be described as [46, 47]:

$$\begin{aligned}
\langle c_j^{in}(t)c_j^{in+}(t)\rangle &= (\mathcal{N}+1)\delta(t-t), \\
\langle c_j^{in+}(t)c_j^{in}(t)\rangle &= \mathcal{N}\delta(t-t), \quad (j=1,2) \\
\langle c_j^{in}(t)c_k^{in}(t)\rangle &= \mathcal{M}e^{-i(\Delta_{c_j}t+\Delta_{c_k}t)}\delta(t-t), \\
\langle c_j^{in+}(t)c_k^{in+}(t)\rangle &= \mathcal{M}^*e^{i(\Delta_{c_j}t+\Delta_{c_k}t)}\delta(t-t), \quad (j\neq k=1,2)
\end{aligned} \quad (4)$$

where N=sinh²r, M=sinhrcoshr. r denotes the squeezing parameter of two-mode squeezed light field. Since the driving frequency and the resonant frequency of optical cavity are not equal, i.e., $\Delta_{cj}, \Delta_{ck} \neq 0$, phase factors related to $\Delta_{cj}$ and $\Delta_{ck}$ are not zero in Eq. (4). The strong microwave (optical) driving field can significantly strengthen the effective coupling rate of the magnon-phonon (phonon-optical cavity) mode. The magnon mode and optical mode have large amplitude $|\langle m_j\rangle|\gg 1$ and $|\langle c_j\rangle|\gg 1$ [29]. Considering the linearly coupling relation of the microwave cavity mode and the magnon mode, the large amplitudes $|\langle a_j\rangle|\gg 1$ is also obtained. Therefore, the dynamics of the system can be linearized and each operator can be described in the form of a fluctuation around a large average value, i.e., O_j = (O_j ) + δO_j (O = a, m, b, c). Neglect the high-order fluctuation terms in the linearization process, the steady-state values of each mode can be obtained from Eq. (2) as follows:

$$\begin{aligned}
\langle a_j\rangle &= \frac{-ig_{a_j}\langle m_j\rangle}{i\Delta_{a_j}+\kappa_{a_j}}, \quad \langle m_j\rangle = \frac{\Omega_j(\kappa_{a_j}+i\Delta_{a_j})}{g_{a_j}^2+(\kappa_{m_j}+i\widetilde{\Delta}_{m_j})(\kappa_{a_j}+i\Delta_{a_j})}, \\
\langle b_j\rangle &= \frac{-ig_{m_j}|\langle m_j\rangle|^2+ig_{c_j}|\langle c_j\rangle|^2}{i\omega_{b_j}+\gamma_j}, \quad \langle c_j\rangle = \frac{E_j}{\kappa_{c_j}+i\widetilde{\Delta}_{c_j}}.
\end{aligned} \quad (5)$$

Here $\widetilde{\Delta}_{mj}=\Delta_{mj}+2g_{mj}\text{Re}(b_j)$ and $\widetilde{\Delta}_{cj}=\Delta_{cj}+2g_{cj}\text{Re}(b_j)$, and both of them contain a detuning term and a frequency shift term. The frequency shift terms are caused by mechanical displacements due to magnetostrictive interaction and radiation pressure interaction, respectively. The frequency shift terms are usually small, $|\widetilde{\Delta}_{mj}|\simeq|\Delta_{mj}|$, $|\widetilde{\Delta}_{cj}|\simeq|\Delta_{cj}|$. $|\Delta_{aj}|,|\widetilde{\Delta}_{mj}|,|\widetilde{\Delta}_{cj}|\simeq\omega_{bj}\gg\kappa_{mj},\kappa_{cj},\gamma_j$. Therefore, the average value of each mode can be safely approximated as $\langle a_j\rangle\simeq-ig_{aj}\langle m_j\rangle/i\Delta_{aj}$, $\langle m_j\rangle\simeq-i\Omega_j\Delta_{aj}/(g^2-\widetilde{\Delta}_{mj}\Delta_{aj})$, $\langle b_j\rangle\simeq-ig_{mj}|\langle m_j\rangle|^2+ig_{cj}|\langle c_j\rangle|^2/i\omega_{bj}$, $\langle c_j\rangle\simeq E_j/i\Delta_{cj}$. The fluctuation terms of the quantum Langevin equations for the system can be further described as:

$$\begin{aligned}
\delta\dot{a}_j &= -(i\Delta_{a_j}+\kappa_{a_j})\delta a_j-ig_{a_j}\delta m_j+\sqrt{2\kappa_{a_j}}a_j^{in}, \\
\delta\dot{m}_j &= -(i\Delta_{m_j}+\kappa_{m_j})\delta m_j-ig_{a_j}\delta a_j-G_{mb_j}(\delta b_j^++\delta b_j)+\sqrt{2\kappa_{m_j}}m_j^{in}, \\
\delta\dot{b}_j &= -(i\omega_{b_j}+\gamma_j)\delta b_j-G_{mb_j}(\delta m_j^+-\delta m_j)+G_{bc_j}(\delta c_j^+-\delta c_j)+\sqrt{2\gamma_j}b_j^{in}, \\
\delta\dot{c}_j &= -(i\Delta_{c_j}+\kappa_{c_j})\delta c_j+G_{bc_j}(\delta b_j^++\delta b_j)+\sqrt{2\kappa_{c_j}}c_j^{in},
\end{aligned} \quad (6)$$

where $G_{mbj} = ig_m \langle m_j \rangle$ ($G_{bcj} = ig_c \langle c_j \rangle$) is the effective magnomechanical (optomechanical) coupling rate. Set $\delta a_j = \delta\tilde{a}_j e^{-i\Delta_{aj}t}$, $\delta m_j = \delta\tilde{m}_j e^{-i\Delta_{mj}t}$, $\delta b_j = \delta\tilde{b}_j e^{-i\Delta_{bj}t}$ and $\delta c_j = \delta\tilde{c}_j e^{-i\Delta_{cj}t}$. The fluctuation terms of the quantum Langevin equations in Eq. (6) can be re-written as [29]:

$$\begin{aligned}
\dot{\delta\tilde{a}}_j &= -\kappa_{a_j}\delta\tilde{a}_j - ig_{a_j}\delta\tilde{m}_j + \sqrt{2\kappa_{a_j}}\tilde{a}_j^{in}, \\
\dot{\delta\tilde{m}}_j &= -\kappa_{m_j}\delta\tilde{m}_j - ig_{a_j}\delta\tilde{a}_j - G_{mb_j}\delta\tilde{b}_j + \sqrt{2\kappa_{m_j}}\tilde{m}_j^{in}, \\
\dot{\delta\tilde{b}}_j &= -\gamma_j\delta\tilde{b}_j + G_{mb_j}\delta\tilde{m}_j - G_{bc_j}\delta\tilde{c}_j + \sqrt{2\gamma_j}\tilde{b}_j^{in}, \\
\dot{\delta\tilde{c}}_j &= -\kappa_{c_j}\delta\tilde{c}_j + G_{bc_j}\delta\tilde{b}_j + \sqrt{2\kappa_{c_j}}\tilde{c}_j^{in}.
\end{aligned} \quad (7)$$

Generally, the quadrature operators are defined as $\delta X_{a_j} = (\delta\tilde{a}_j + \delta\tilde{a}_j^+)/\sqrt{2}$, $\delta Y_{a_j} = i(\delta\tilde{a}_j - \delta\tilde{a}_j^+)/\sqrt{2}$, $\delta X_{m_j} = (\delta\tilde{m}_j + \delta\tilde{m}_j^+)/\sqrt{2}$, $\delta Y_{m_j} = i(\delta\tilde{m}_j - \delta\tilde{m}_j^+)/\sqrt{2}$, $\delta q_j = (\delta\tilde{b}_j + \delta\tilde{b}_j^+)/\sqrt{2}$, $\delta p_j = i(\delta\tilde{b}_j - \delta\tilde{b}_j^+)/\sqrt{2}$, $\delta X_{c_j} = (\delta\tilde{c}_j + \delta\tilde{c}_j^+)/\sqrt{2}$, $\delta X_{c_j} = i(\delta\tilde{c}_j - \delta\tilde{c}_j^+)/\sqrt{2}$ [21, 29]. The fluctuating quadrature terms of the quantum Langevin equations can be described as a matrix form:

$$\dot{u}(t) = Au(t) + n(t), \quad (8)$$

where $u(t) = (\delta X_{a_1}, \delta Y_{a_1}, \delta X_{a_2}, \delta Y_{a_2}, \delta X_{m_1}, \delta Y_{m_1}, \delta X_{m_2}, \delta Y_{m_2}, \delta q_1, \delta p_1, \delta q_2, \delta p_2, \delta X_{c_1}, \delta Y_{c_1}, \delta X_{c_2}, \delta Y_{c_2},)^T$, $n(t) = (\sqrt{2\kappa_{a_1}}\delta X_{a_1}^{in}, \sqrt{2\kappa_{a_1}}\delta Y_{a_1}^{in}, \sqrt{2\kappa_{a_2}}\delta X_{a_2}^{in}, \sqrt{2\kappa_{a_2}}\delta Y_{a_2}^{in}, \sqrt{2\kappa_{m_1}}\delta X_{m_1}^{in}, \sqrt{2\kappa_{m_1}}\delta Y_{m_1}^{in}, \sqrt{2\kappa_{m_2}}\delta X_{m_2}^{in}, \sqrt{2\kappa_{m_2}}\delta Y_{m_2}^{in}, \sqrt{2\gamma_1}\delta q_1^{in}, \sqrt{2\gamma_1}\delta p_1^{in}, \sqrt{2\gamma_2}\delta q_2^{in}, \sqrt{2\gamma_2}\delta p_2^{in}, \sqrt{2\kappa_{c_1}}\delta X_{c_1}^{in}, \sqrt{2\kappa_{c_1}}\delta Y_{c_1}^{in}, \sqrt{2\kappa_{c_2}}\delta X_{c_2}^{in}, \sqrt{2\kappa_{c_2}}\delta Y_{c_2}^{in})$. $A$ is the drift matrix:

$$A = \begin{pmatrix} A_a & A_{am} & A_{ab} & A_{ac} \\ A_{am} & A_m & -A_{mb} & A_{mc} \\ A_{ab} & A_{mb} & A_b & -A_{bc} \\ A_{ac} & A_{mc} & A_{bc} & A_c \end{pmatrix}, \quad (9)$$

where $A_o = -\text{diag}(\kappa_{o_1}, \kappa_{o_1}, \kappa_{o_2}, \kappa_{o_2})$, $(o = a, m, c)$, $A_b = -\text{diag}(\gamma_1, \gamma_1, \gamma_2, \gamma_2)$. $A_{mb} = \text{diag}(G_{mb_1}, G_{mb_1}, G_{mb_2}, G_{mb_2})$ and $A_{bc} = \text{diag}(G_{bc_1}, G_{bc_1}, G_{bc_2}, G_{bc_2})$ denote the magnon-phonon and phonon-optical cavity coupling matrices. $A_{ab}, A_{ac}$ and $A_{mc}$ are the 4×4 zero matrices. The coupling matrix of the magnon-microwave cavity is $A_{am} = (0, g_{a_1}, 0, 0; -g_{a_1}, 0, 0, 0; 0, 0, 0, -g_{a_2}; 0, 0, -g_{a_2}, 0)$.

To prepare the stationary entanglement of microwave fields, all eigenvalues of the drift matrix A must be negative real number due to the Routh-Hurwitz criterion [48]. For the linearized dynamics and Gaussian input noises, the system still preserves Gaussian states. The steady state of the system is an eight-mode Gaussian state, and it can be fully characterized by a 16×16 covariance matrix (CM) [49]. $C_{ij}(t,t') = \frac{1}{2}\langle u_i(t)u_j(t') + u_j(t')u_i(t)\rangle$, (i, j=1,2,...,16). The steady-state CM can be obtained by solving the Lyapunov equation [50]:

$$AV + VA^T = -D, \quad (10)$$

Here $D = D_a \oplus D_m \oplus D_b \oplus D_c$ is the diffusion matrix, and it is defined as $D_{ij}\delta(t-t') = \frac{1}{2}\langle n_i(t)n_j(t') + n_j(t')n_i(t)\rangle$, $D_a = \text{diag}[\kappa_{a_1}(2N_{a_1}+1), \kappa_{a_1}(2N_{a_1}+1), \kappa_{a_2}(2N_{a_2}+1), \kappa_{a_2}(2N_{a_2}+1)]$, $D_m = \text{diag}[\kappa_{m_1}(2N_{m_1}+1), \kappa_{m_1}(2N_{m_1}+1), \kappa_{m_2}(2N_{m_2}+1), \kappa_{m_2}(2N_{m_2}+1)]$, $D_b = \text{diag}[\gamma_1(2N_{b_1}+1), \gamma_1(2N_{b_1}+1), \gamma_2(2N_{b_2}+1), \gamma_2(2N_{b_2}+1)]$. $D_c$ is the input noise diffusion matrix of the squeezed light field, and it is related to two optical cavity modes. $D_c$ can be described as:

$$D_c = \begin{pmatrix} \kappa_{c_1}(2\mathcal{N}+1) & 0 & \sqrt{\kappa}(\mathcal{M}+\mathcal{M}^*) & i\sqrt{\kappa}(-\mathcal{M}+\mathcal{M}^*) \\ 0 & \kappa_{c_1}(2\mathcal{N}+1) & i\sqrt{\kappa}(-\mathcal{M}+\mathcal{M}^*) & -\sqrt{\kappa}(\mathcal{M}+\mathcal{M}^*) \\ \sqrt{\kappa}(\mathcal{M}+\mathcal{M}^*) & i\sqrt{\kappa}(-\mathcal{M}+\mathcal{M}^*) & \kappa_{c_2}(2\mathcal{N}+1) & 0 \\ i\sqrt{\kappa}(-\mathcal{M}+\mathcal{M}^*) & -\sqrt{\kappa}(\mathcal{M}+\mathcal{M}^*) & 0 & \kappa_{c_2}(2\mathcal{N}+1) \end{pmatrix}, \quad (11)$$

where $\kappa = \kappa_{c_1}\kappa_{c_2}$. By solving the Lyapunov equation of Eq. (9) and Lyapunov equation of Eq. (10), the logarithmic negativity $E_N$ can be obtained which quantifies the bipartite entanglement characteristic. $E_N$ is generally defined as [44, 51]:

$$E_N = \max\left[0, -\ln 2\eta^-\right] \quad (12)$$

where $\eta^- = 2^{-1/2}\left[\Sigma - \left(\Sigma^2 - 4\det V_4\right)^{1/2}\right]^{1/2}$, $\Sigma = \det V_{o_1} + \det V_{o_2} - 2\det V_{o_1 o_2}$, (O =a, m, b, c). $V_{O1}$ and $V_{O2}$ are 2×2 block matrices which denote coefficients corresponding to $O_1$ and $o_2$ modes in CM. $V_{O1O2}$ denotes the correlation terms of $O_1$ and $O_2$ modes in CM. $V_4 = \left[V_{o_1}, V_{o_1 o_2}; V_{o_1 o_2}^T, V_{o_2}\right]$ is a 4×4 matrix. Two Gaussian mode light fields are entangled when $\eta^- < \frac{1}{2}$ is satisfied [48, 49].

### 3. Numerical results of microwave entanglement

Fig. 2 shows stationary entanglements of two optical cavity modes $E_{c1c2}$, phonon modes $E_{b1b2}$, magnon modes $E_{m1m2}$ and microwave cavity modes $E_{a1a2}$, respectively. For simplicity, we assume that two cavity opto-magnomechanical subsystems are symmetrical. The parameters used in our scheme are feasible in many relevant experiments as $\omega_{aj} = \omega_{mj} = 2\pi \times 10$ GHz, (j = 1, 2), $\omega_{bj} = 2\pi \times 40$ MHz, $\kappa_{aj} = 2\pi \times 1.5$ MHz, $\kappa_{mj} = 1/5\kappa_{aj}$, $\kappa_{cj} = 2\pi \times 3$ MHz, $\gamma_{bj} = 2\pi \times 100$ Hz, $g_{aj} = 2\pi \times 4$ MHz and T = 10 mK [21, 28, 30]. $G_{mbj}$ is determined by the driving magnetic field intensity $H_{d_j} = \sqrt{2\mu_0 P_j / (l_j \omega_j c)}$ and power $P_j = 0.91$ mW, where $\mu_0 = 4\pi \times 10^{-7}$ is the vacuum magnetic permeability, $l_j$ ($w_j$) is the length (width) of the YIG micro-bridge [30]. In Fig. 2(a), all beam splitter interactions are not activated when the effective coupling rate satisfies $G_{mb} < \kappa_a$ or $G_{bc} < 0.4\kappa_c$. So the density distribution of two optical cavity modes entanglement $E_{c1c2}$ is close to two coordinate axes. The effective coupling rate $G_{mbj} = ig_m(m_j)$ ($G_{bcj} = ig_c(c_j)$) increases when the driving microwave (optical) fields are strong red detuning. The entanglement of two optical fields transfers to two microwave fields (red area in Fig. 2(d)) when the effective coupling rates $G_{mb}$ and $G_{bc}$ increases (blue area in Fig.2(a)). Fig. 2(b) shows a stationary entanglement of two phonon modes. The strong red detuned driving optical fields activate the optomechanical beam splitter interactions. When the effective coupling rate $G_{mb}$ is fixed, the entanglement $E_{b1b2}$ increases as the effective coupling rate $G_{bc}$ increases. The optomechanical beam splitter interactions are invalid when the effective coupling rate $G_{bc} = 0$, which leads to the disappearance of entanglement ($E_{b1b2} = 0$). Fig. 2(c) shows the entanglement of two magnon modes. Both optomechanical and magnomechanical beam splitter interactions are activated due to the strong red-detuned microwave and optical fields injected from two ends of the system. The entanglement $E_{c1c2}$ increases when the effective coupling rates $G_{mb}$ or $G_{bc}$ increases. Fig. 2(d) shows the entanglement of two microwave fields. As the effective coupling rates $G_{mb}$ or $G_{bc}$ increases, it clearly shows that the entanglement is transferred from magnon modes to microwave cavity modes via the magnetic dipole interactions. A optimal microwave fields entanglement $E_{a1a2} = 0.54$ is obtain when $G_{bcj} = 2\pi \times 10$ MHz and $G_{mbj} = 2\pi \times 4.5$ MHz. The quantum correlation is efficiently transferred from optical modes to phonon modes, then to magnon modes, and finally to microwave modes via the interactions of optomechanical, magnomechanical, and magnon-microwave beam splitters. The transfer efficiency of the entanglement is 26.5%.

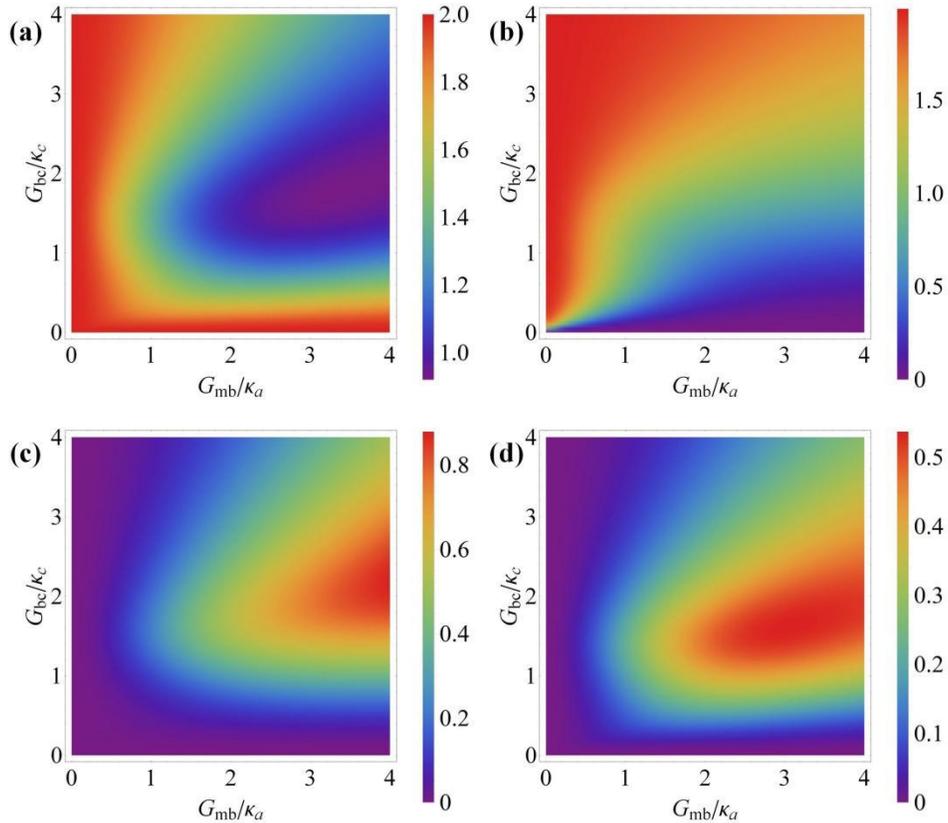

Figure. 2. Steady-state density diagram of bipartite entanglement. The entanglement of (a) two optical cavity modes $E_{c1c2}$, (b) two phonon modes $E_{b1b2}$, (c) two magnon modes $E_{m1m2}$ and (d) two microwave cavity modes $E_{a1a2}$ versus the effective coupling rates $G_{mb}$ and $G_{bc}$ with a two-mode squeezed optical driving field (r=1). The colored columns at the right sides of the pictures indicate different entanglement strengths of corresponding cavity modes.

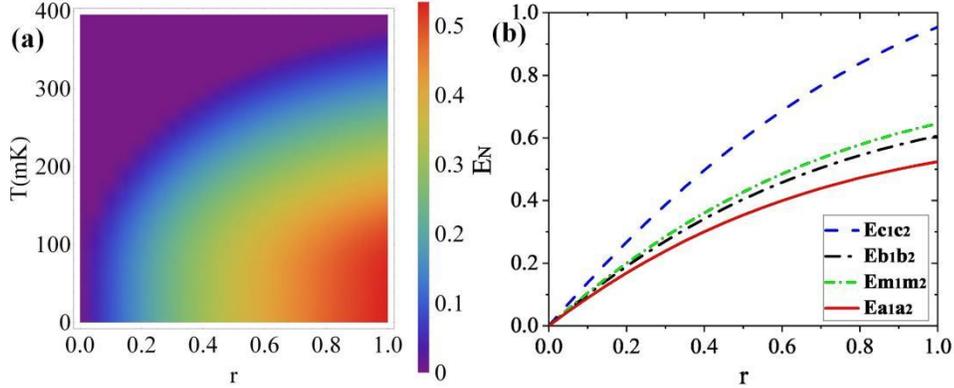

Figure. 3. Bipartite entanglement versus temperature T (mK) and squeezing parameter r. (a) The entanglement $E_{a_1a_2}$ of two microwave cavity modes versus the squeezing parameter r and temperature T. The colored column at the right side indicates the entanglement strength. (b) The variation of stationary entanglement $E_{c_1c_2}$, $E_{b_1b_2}$, $E_{m_1m_2}$ and $E_{a_1a_2}$ versus the squeezing parameter r with T = 10 mK. $G_{mb} = 2.8\kappa_a$, $G_{bc} = 1.6\kappa_c$. Other parameters are same as Fig.2.

Next, we show that the entanglement of microwave fields is robust to thermal bath noise. In Fig. 3(a), the entanglement $E_{a_1a_2}$ still exists when the environmental temperature reaches T=385 mK and the squeezing parameter is r=1. It means that the robustness of entanglement $E_{a_1a_2}$ strengthens as squeezing parameter r increases. Fig. 3(b) shows the variation of entanglement $E_N$ versus the squeezing parameter r under the steady-state condition. The bipartite entanglement can be simultaneously distributed in four different entangled states $E_{c_1c_2}$, $E_{b_1b_2}$, $E_{m_1m_2}$ and $E_{a_1a_2}$ when the cavities are continuously driven. Meanwhile, the entanglement $E_N$ increases as the squeezing parameter r increases.

Fig. 4(a-c) shows the matching of effective coupling rates for magnon-microwave $g_{aj}$, magnon-phonon $G_{mbj}$, phonon-optical cavity $G_{bcj}$ in two subsystems. There is a wide tunable range to match effective coupling rate $G_j$ (G =$g_a$, $G_{mb}$, $G_{bc}$) (j=1, 2) of two subsystems when the entanglement is small. However, the effective coupling rates $G_1$ and $G_2$ in the two cavity opto-magnomechanical subsystems should be as consistent as possible in order to obtain large entanglement. Therefore, two cavity opto-magnomechanical subsystems are easier to be matched when the squeezing parameter r=0.4, in which the effective coupling rates $G_1$ and $G_2$ have a wide tunable range to be consistent. With the increasing of squeezing parameter r, the matching of two subsystems becomes difficult while the entanglement $E_N$ increases. Fig. 4(d) shows a density diagram of microwave field entanglement versus the coupling rate $g_a$ and the squeezing parameter r. The coupling of the magnon-microwave cavity has a efficient response for entanglement when the value of $g_a$ is in the range of $2.5\kappa_a \sim 4\kappa_a$.

We note our scheme is different from the scheme in Ref. [21] both in principle and result. The entanglement of Ref. [21] originates from the nonlinear magnetostrictive effect and two microwave fields are injected in a same cavity, which can be explained by a cross-shaped cavity model. As a result, the entanglement EN of two microwave fields is a fixed value, and it can not be adjusted freely in this case. In contrast, in our scheme the entanglement originates from the two-mode squeezed optical field and two microwave fields are injected into two cavity opto-magnomechanical subsystems, respectively. The entanglement is transferred from the optical cavities to microwave cavities via beam splitter interactions. Therefore, the strength of the microwave fields entanglement can be adjusted freely in a wide range as the squeezing parameter r changes. Furthermore, the proposed scheme has higher entanglement degree and temperature robustness comparing to Ref. [21].

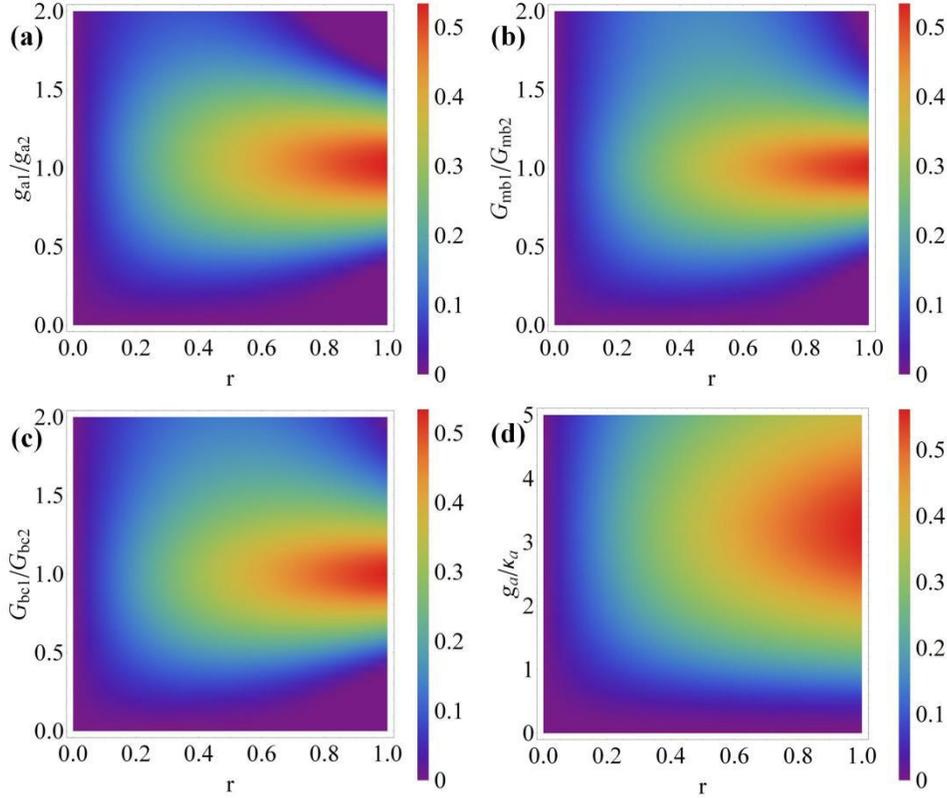

Figure. 4. Bipartite entanglement diagram versus the effective coupling rate and the squeezing parameter r. (a), (b) and (c) show the variation of stationary entanglement $E_{a1a2}$ versus the squeezing parameter r and three different effective coupling rates $g_{aj}$, $G_{mbj}$ and $G_{bcj}$ (j = 1, 2), respectively. (d) Steady-state density of entanglement $E_{a1a2}$ versus the squeezing parameter r and the coupling rate $g_a/\kappa_a$ of magnon-microwave cavity. $G_{mb} = 2.8\kappa_a$, $G_{bc} = 1.6\kappa_c$. The colored columns at the right sides of the pictures indicate the corresponding entanglement strength. Other parameters are same as Fig.2.

## 4. Conclusion

In summary, we demonstrated a scheme to generate high-efficiency and robust entanglement of two microwave fields in a dual opto-magnomechanical system. The entangled state of two optical fields is transferred to two microwave fields via state- swap interactions. It demonstrates that the prepared entanglement of two microwave fields is high-efficiency ($E_{a1a2}$ = 0.54) and robust (T=385 mK) in the dual-cavity opto-magnomechanical system with the squeezing parameter of injected filed r = 1. All parameters used in our scheme are feasible in many practical experiments [28, 37, 52]. Therefore, it is expected to be successfully implemented in experiment. Due to the low absorption rate and strong diffraction effect, high-efficiency entanglement of microwave fields is usually difficult to be realized [53]. Therefore, the proposed scheme paves the way for future applications of entangled-state microwave fields. Our result has various applications which require high and robust entanglement of microwave fields such as quantum radar [40, 41], microwave quantum illumination [54, 55], quantum communication in free space [56], quantum wireless fidelity (Wi-Fi) network, etc.

**Funding**

This work was supported by National Natural Science Foundation of China (Grant Nos. 11704053, 52175531); the Science and Technology Research Program of Chongqing Municipal Education Commission (Grant No. KJQN201800629); the Postdoctoral Applied Research Program of Qingdao (Grant No. 62350079311135); the Postdoctoral Applied Innovation Program of Shandong (Grant No. 62350070311227).

**Disclosures**

The authors declare that there are no conflicts of interest related to this article.